\documentclass[twocolumn]{aastex62}
\usepackage{epstopdf}
\usepackage{wrapfig}
\usepackage{lipsum} 
\usepackage[utf8]{inputenc}
\usepackage{amsmath}
\usepackage{natbib}
\bibpunct{(}{)}{;}{a}{}{,}

\newcommand*\shellatm{\textsc{Shell-Atm}}

\usepackage{color}
\definecolor{darkgreen}{RGB}{0,142,128}
\definecolor{darkblue}{RGB}{0,100,170}
\usepackage{hyperref}
\hypersetup{colorlinks,citecolor=blue,linkcolor=blue}

\graphicspath{{./}{figures/}}

\submitjournal{Frontiers in Astronomy and Space Sciences}

\shorttitle{FIP effect and turbulence in open and closed field region}
\shortauthors{Réville et al.}
\begin{document}

\title{Investigating the origin of the FIP effect with a shell turbulence model}

\author[0000-0002-2916-3837]{Victor Réville}
%\email{victor.reville@irap.omp.eu}
\affiliation{IRAP, Universit\'e Toulouse III - Paul Sabatier,
CNRS, CNES, Toulouse, France}

\author[0000-0003-4039-5767]{Alexis P. Rouillard}
\affiliation{IRAP, Universit\'e Toulouse III - Paul Sabatier,
CNRS, CNES, Toulouse, France}

\author[0000-0002-2381-3106]{Marco Velli}
\affil{UCLA Earth Planetary and Space Sciences Department, LA, CA, USA}

\author[0000-0003-4380-4837]{Andrea Verdini}
\affiliation{Dipatimento di Fisica e Astronomia, Universit\'a di Firenze, Sesto Fiorentino, Italia}

\author[0000-0003-4290-1897]{Éric Buchlin}
\affiliation{Université Paris-Saclay, CNRS,  Institut d'Astrophysique Spatiale, 91405, Orsay, France}

\author[0000-0001-6216-6530]{Michaël Lavarra}
\affiliation{IRAP, Universit\'e Toulouse III - Paul Sabatier,
CNRS, CNES, Toulouse, France}

\author[0000-0002-1814-4673]{Nicolas Poirier}
\affiliation{IRAP, Universit\'e Toulouse III - Paul Sabatier,
CNRS, CNES, Toulouse, France}

\begin{abstract}
The enrichment of coronal loops and the slow solar wind with elements that have low First Ionisation Potential, known as the FIP effect, has often been interpreted as the tracer of a common origin. A current explanation for this FIP fractionation rests on the influence of ponderomotive forces and turbulent mixing acting at the top of the chromosphere. The implied wave transport and turbulence mechanisms are also key to wave-driven coronal heating and solar wind acceleration models. This work makes use of a shell turbulence model run on open and closed magnetic field lines of the solar corona to investigate with a unified approach the influence of magnetic topology, turbulence amplitude and dissipation on the FIP fractionation. We try in particular to assess whether there is a clear distinction between the FIP effect on closed and open field regions.
\end{abstract}


\section{Introduction} 
\label{intro}

The First Ionization Potential (FIP) effect is an enrichment of heavy elements with low-FIP such as Fe, Si and Mg compared with photospheric abundances. It was initially measured in the solar wind and Solar Energetic Particles (SEPs) and later inferred from spectroscopic observations of the corona \citep[see][and references therein]{Meyer1985a,Meyer1985b,Bochsler1986,Gloeckler1989,Feldman1992}. The FIP bias, i.e. the ratio of coronal to photospheric abundances, is moreover mass independent. This means that processes below the transition region are strongly affecting the hydrostatic balance of the partially ionized chromosphere. Early on, explanations of the FIP effect involved diffusion, flows or some sort of turbulent mixing in the chromosphere to prevent any gravitational settling \citep[see, e.g.][]{Marsch1995,Schwadron1999}.

\citet{Schwadron1999} favored the hypothesis of turbulent wave heating in the chromosphere as a way to both prevent a mass-dependant fractionation and to obtain a low-FIP bias. Later on, \citet{Laming2004} proposed the ponderomotive acceleration, i.e. the time averaged gradient of magnetic fluctuations, as the origin of the FIP effect. The ponderomotive acceleration has this advantage that it may change sign and could explain the inverse FIP effect observed in low-mass stars \citep{WoodLinsky2010,Laming2015}. Both processes rely on Alfvén waves propagating parallel and anti-parallel to the magnetic field, to trigger a turbulent cascade through non-linear interactions and heating. These wave populations naturally arise in coronal loops, where footpoints motions excite the loop on both ends, but they are also expected in the open solar wind where reflection on large scale gradients \citep{Velli1989,ZhouMatthaeus1989} or compressible instabilities \citep{TeneraniVelli2013,Shoda2018b,Reville2018} create an sunward component from a purely anti-sunward wave packet.

The Ulysses spacecraft have shown clear composition differences between the fast and the slow wind components, the slow wind showing FIP biases close to the one observed in coronal loops \citep{Geiss1995}. After all, decades of observations \citep{BelcherDavis1971,TuMarsch1995,BrunoCarbone2013} and modeling of fluctuations in the solar wind have now brought compelling evidence that turbulence is likely to be a fundamental ingredient of coronal heating and solar wind acceleration \citep[see, e.g.,][]{VerdiniVelli2007,PerezChandran2013,Shoda2018b,Reville2020ApJS}. Perturbations, or waves are also essential to provide the additional acceleration giving birth to the fast wind ($\geq 600$\,km/s), through the ponderomotive force \citep{AlazrakiCouturier1971,Belcher1971,Jacques1977,Leer1982}. The low-FIP bias of the slow wind is often understood as the proof that the it originates in coronal loops, through exchange and magnetic reconfiguration (or reconnection) in the low corona \citep[see, e.g.,][]{Antiochos2012}. Yet, if turbulence is both responsible for the solar wind acceleration and the FIP effect, can we rule out a scenario where FIP fractionation occurs in purely open regions?

We investigate this question using a coupled modelling approach. First, we extract unidimensional profiles of open solar wind flux tubes and coronal loops using a multi-dimensional MHD code \citep{Reville2020ApJS}. We then use the \shellatm\ code \citep{BuchlinVelli2007,Verdini2009} to compute the propagation, cascade and dissipation of purely transverse perturbations of velocity and magnetic field along these profiles.  We perform a parameter study where we vary the geometry, the initial perturbation amplitude and the initial injection scale, and discuss their effect on the turbulence properties in the chromosphere and transition region. We estimate the resulting FIP biases with an analytical fractionation model. 

\section{Background wind and transition region profiles}
\label{sec:tr}

In this section, we describe the global MHD simulation used to extract the different wind profiles later input in our turbulence model. The code itself is a global MHD solver based on PLUTO \citep{Mignone2007} and our current implementation is described in details in \citet{Reville2020ApJS} and \citet{Reville2020ApJL}. The main purpose of the global simulation is to provide a realistic structure of the transition region, which plays an essential part in the present work. We rely on a single run, of a dipolar solar field of 5 G, which is typical of solar minimum configurations \citep[see, e.g.,][]{DeRosa2012}, and a uniform coronal heating prescribed with the following function: 
\begin{equation}
Q_h = F_h/H \left(\frac{R_{\odot}}{r} \right)^2 \exp{ \left(-\frac{r-R_{\odot}}{H}\right)},
\label{eq:adhoc}
\end{equation}
where $H=1 R_{\odot}$, and $F_h = 10^5$ erg.cm$^{-2}$s$^{-1}$. The value of $F_h$ corresponds to the global energy output of the solar wind.

\begin{figure}
\center
\includegraphics[width=3.3in]{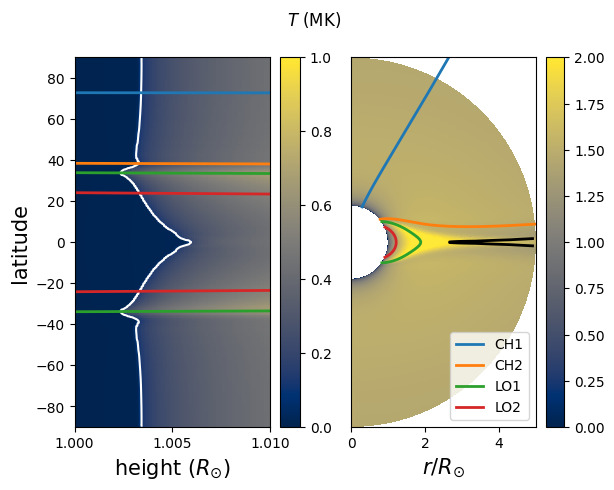}
\caption{Plasma temperature obtained with the global MHD model. On the left panel, we show the temperature profile as a function of latitude and height. The white line shows the contour $T=10^5$ K, characterizing the location of the transition region. Blue lines and orange lines are open field lines (referred as CH for coronal holes), green and red lines are coronal loops (referred as LO). They are almost perfectly radial at this scale. On the right panel, we show an extended view of the solution and integrated field lines in the meridional plane up to $5R_{\odot} $. The black line on the right is the Alfvén surface.}
\label{fig:transition}
\end{figure}

The result of the simulation is shown in Figure \ref{fig:transition}. The left panel is a zoom on the transition region as a function of height and latitude. Contrarily with previous studies \citep{Reville2020ApJS,Reville2020ApJL}, whose inner boundary was located at the base of the corona, we start the simulation at the photosphere with a temperature of $6000$ K. We initialize the simulation from a one dimensional profile of the solar atmosphere obtained with the 1D version of the code described in \citet{Reville2018} and the boundary conditions are identical to \citet{Reville2020ApJS}. We use a spherical grid with a radial resolution that goes from $3 \times 10^{-5} R_{\odot}$ to $0.3 R_{\odot}$ from the photosphere to 20 $R_{\odot}$. There are 544 points in the radial and 512 in the latitudinal direction. The location of the transition region (TR) varies with the latitude and the temperature profiles. It is relatively constant in coronal holes  at around $1.003 R_{\odot}$, i.e. $2000$ km above the photosphere. The minimum height corresponds to the highest coronal temperatures at the very edge of the helmet streamer. The height of the TR then increases within the core of the streamer up to $1.007 R_{\odot}$, accordingly with lower coronal temperatures. These lower temperatures are likely due to increased density that result in increased cooling inside the loop \citep[radiative losses are proportional to the squared density in the model, see][]{Reville2020ApJS}. This structure has already been described in the work of \citet{Lionello2001} using a similar heating function.

The shock capturing ability of PLUTO, based on Riemann solvers, is essential to describe strong density and temperature gradients defining the transition region. \citet{Reville2020ApJS} have developed an additional module to propagate Alfvén waves in the corona. Yet, since this module is based on a Wentzel-Krimers-Brillouin approximation, it is not suited for a precise description of the transition region. To further study the role of turbulence in the composition of the solar wind, we thus rely on the \shellatm\ code, which solves the non-linear incompressible Alfvén wave equations on given profiles along magnetic fields lines, neglecting magneto-acoustic modes. We extracted four different solar wind profiles from our 2.5D simulation, shown in Figure \ref{fig:transition}. In the following section, we describe the \shellatm\ simulations made from these profiles. 

\section{Shell turbulence model}
\label{sec:shell}

\begin{figure}
\center
\includegraphics[width=3.3in]{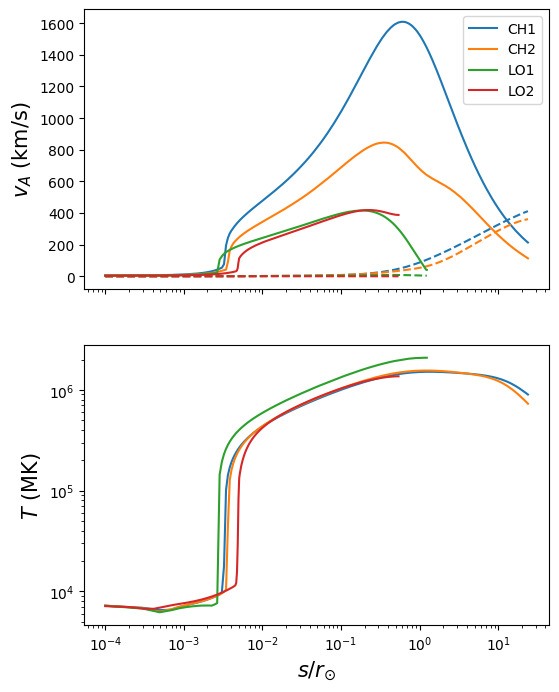} \\
\caption{Profiles of the Alfvén speed, flow speed (in dashed lines) and temperature for the four field lines extracted from the simulation, along the curvilinear abscissa $s$. For loops, we only show one side up to the apex. These profiles will be the source of the amplification and reflection (particularly in open regions) of perturbations, the latter leading to the creation of counter-propagating perturbations. Non-linear interactions between counter-propagating waves create the cascade and the dissipation at small scales.}
\label{fig:profiles}
\end{figure}

\shellatm\ is a low dimension fluid turbulence code based on the shell technique \citep[see][]{GiulianiCarbone1998,Nigro2004}, that can model coronal loops \citep{Nigro2004,BuchlinVelli2007} and open solar wind solutions \citep{Verdini2009}. As fixed background profiles of flow speed, density and temperature, we use the four profiles extracted from our global MHD simulation. They are shown in Figure \ref{fig:profiles}. Cases identified with CH correspond to open field regions in Figure \ref{fig:transition}. The CH1 profile is well in the coronal hole of the northern hemisphere in the fast wind region. CH2 is located close to the streamer and thus in the slow wind region of the simulation. LO cases are coronal loops, the LO1 case being at the open/close boundary with the minimum height of the transition region and LO2 being the smallest loop, well inside the helmet streamer.

\shellatm\ solves the reduced (incompressible) evolution of Alfvénic fluctuations, i.e. two coupled equations for the evolution of the parallel and anti-parallel wave populations. The transverse components of the fluctuations are discretized in the spectral space, starting from a scale $k_0$, and then into 21 other shells at scales $k_n=2^n k_0$. We define $z^{\pm}_n = \delta v_n \mp \delta b_n/\sqrt{\mu_0 \rho}$ the Elsässer perturbations at scale $\lambda_n = \lambda_0 2^{-n}$. The equations controlling the transport and dissipation of the $z^{\pm}$ fields are described in details in \citet{BuchlinVelli2007,Verdini2009}. Non-linear interactions are operated in triads following the model of \citet{GiulianiCarbone1998}. Dissipation in \shellatm\ is obtained via explicit resistivity and viscosity. We set the magnetic Prandtl number $\nu/\eta$ to unity and the Lundquist number $S=L v_A/\eta$ lies between $10^6$ in the chromosphere and $10^8$ in the corona. We write the heating per unit mass created by the cascade : 

\begin{equation}
    Q_{\nu}/\rho = \frac{1}{2} \sum_n \nu k_n^2 (|z_n^+|^2 + |z_n^-|^2).
    \label{eq:heating}
\end{equation}

Transverse motions' amplitude $\delta v_{\odot}$ is forced at the base of the domain in the chromosphere over three shells, starting at the scale $k_3 = 8 k_0$. The forcing uses random fluctuation phases that are reset over a correlation time $T_*$ \cite[see][]{BuchlinVelli2007}. This is akin to setting a parallel frequency for the Alfvénic perturbations. $T_*$ is chosen around a few hundreds seconds to follow observations of transverse motions in the corona \citep[see the review of][]{Nakariakov2005}.

We chose to maintain the amplitude $\delta v_{\odot}$ at the lower boundary, around 100 km above the photosphere, and on both boundaries for coronal loops configurations, using the following condition:
\begin{equation}
    z_n^\pm (t) = 2 \delta v_{\odot} - z_n^\mp (t).
\end{equation}
This makes the inner boundary condition partially reflective, with a reflection coefficient being a function of time and of the balance of inward and outward wave populations. Imposing a fully reflecting inner boundary conditions leads indeed to an increase of the base velocity perturbations up to unrealistic values. 

Next, we define
\begin{equation}
    F^\pm = \frac{1}{4} \rho (v \pm v_A)  \sum_n [z^{\pm}_n|^2,
\end{equation}
the energy flux of the perturbations at any given location in the domain. For both coronal holes and loops cases, we have the total energy injected per second and surface unit at a given time

\begin{eqnarray}
    E_{\mathrm{tot}} &=& F_{\mathrm{in}} A_0 - F_{\mathrm{out}} A_1,\\
                     &=& (F^+_0 + F^-_0) A_0 - (F^+_1 + F^-_1) A_1,
\end{eqnarray}
where the subscripts $0$ and $1$ denote the bottom and top of the computational domain respectively. In a statistical steady state, we expect to have the losses in the volume compensating for the input energy, i.e.:

\begin{equation}
\label{eq:balance}
    \langle E_{\mathrm{tot}} \rangle \sim \langle H_\nu \rangle + \langle W \rangle,
\end{equation}
where 
\begin{equation}
    H_\nu = \int Q_\nu  A_{\mathrm{exp}} \; \mathrm{d}s, 
\end{equation}
is the integrated turbulent heating in the domain and $W$ is the work of the force exerted by the perturbations on the solar wind flow:
\begin{equation}
    W = \int \rho \, (\mathbf{u} \cdot \mathbf{a}_w) \, A_{\mathrm{exp}} \; \mathrm{d}s.
\end{equation}

The component of the ponderomotive acceleration $\mathbf{a}_{w}$ along field lines can be written \citep{LitwinRosner1998,Laming2004,Laming2009,Laming2012,Dahlburg2016}:
\begin{equation}
    a_{w,s} = \frac{\partial}{\partial s} \left[\frac{\langle \delta \mathbf{E}^2 \rangle}{2 |B|^2} \right],
    \label{eq:pond}
\end{equation}
where $\langle \delta \mathbf{E} \rangle$ is the time averaged electric field due to the perturbations. Equation \eqref{eq:pond} also assumes that the ion cyclotron frequency is much larger than the wave frequency, which easily verified in the inner heliosphere for typical Alfvén wave spectra and makes the force mass independent. In the remainder of the text, we will refer to equation \eqref{eq:pond} by ponderomotive force (per unit mass and unit volume), or ponderomotive acceleration indifferently.

As we consider one dimensional profile, we use the normalized area $A_{\mathrm{exp}}$, which is unity at the base of all flux tubes, and we write equation~\eqref{eq:balance} in erg.cm$^{-2}$.s$^{-1}$. 

\begin{table}
\caption{Parameters and outputs for the \shellatm\ simulations}
\label{table:shell}
\center
\setlength{\tabcolsep}{3pt}
\begin{tabular}{|c|c|c|c|c|c|}
\hline
Cases & $\delta v_{\odot}$ & $T_*$ (s) & $L^0_{\mathrm{inj}}$ & $\langle H_\nu \rangle$ & $\langle W \rangle$ \\
& (km/s) & (s) & (km) & ($10^5$ cgs) & ($10^5$ cgs) \\
\hline
CH1 A & 7 & 600 & 35000 & 0.84 & 0.49 \\
CH2 A & 7 & 600 & 35000 & 1.8 & 1.0 \\
LO1 A & 7 & 600 & 35000 & 20 & - \\
LO2 A & 7 & 600 & 35000 & 10 & - \\
CH1 B & 10 & 600 & 35000 & 3.1 & 1.6 \\
CH2 B & 10 & 600 & 35000 & 4.6 & 2.9 \\
LO1 B & 10 & 600 & 35000 & 26 & - \\
LO2 B & 10 & 600 & 35000 & 21 & - \\
CH1 C & 3 & 600 & 3500 & 1.2 & 0.081 \\
CH2 C & 3 & 600 & 3500 & 1.2 & 0.092 \\
LO1 C & 3 & 600 & 3500 & 2.6 & - \\
LO2 C & 3 & 600 & 3500 & 2.3 & - \\
\hline
\end{tabular}
\end{table}

\begin{figure*}
\center
\includegraphics[width=7in]{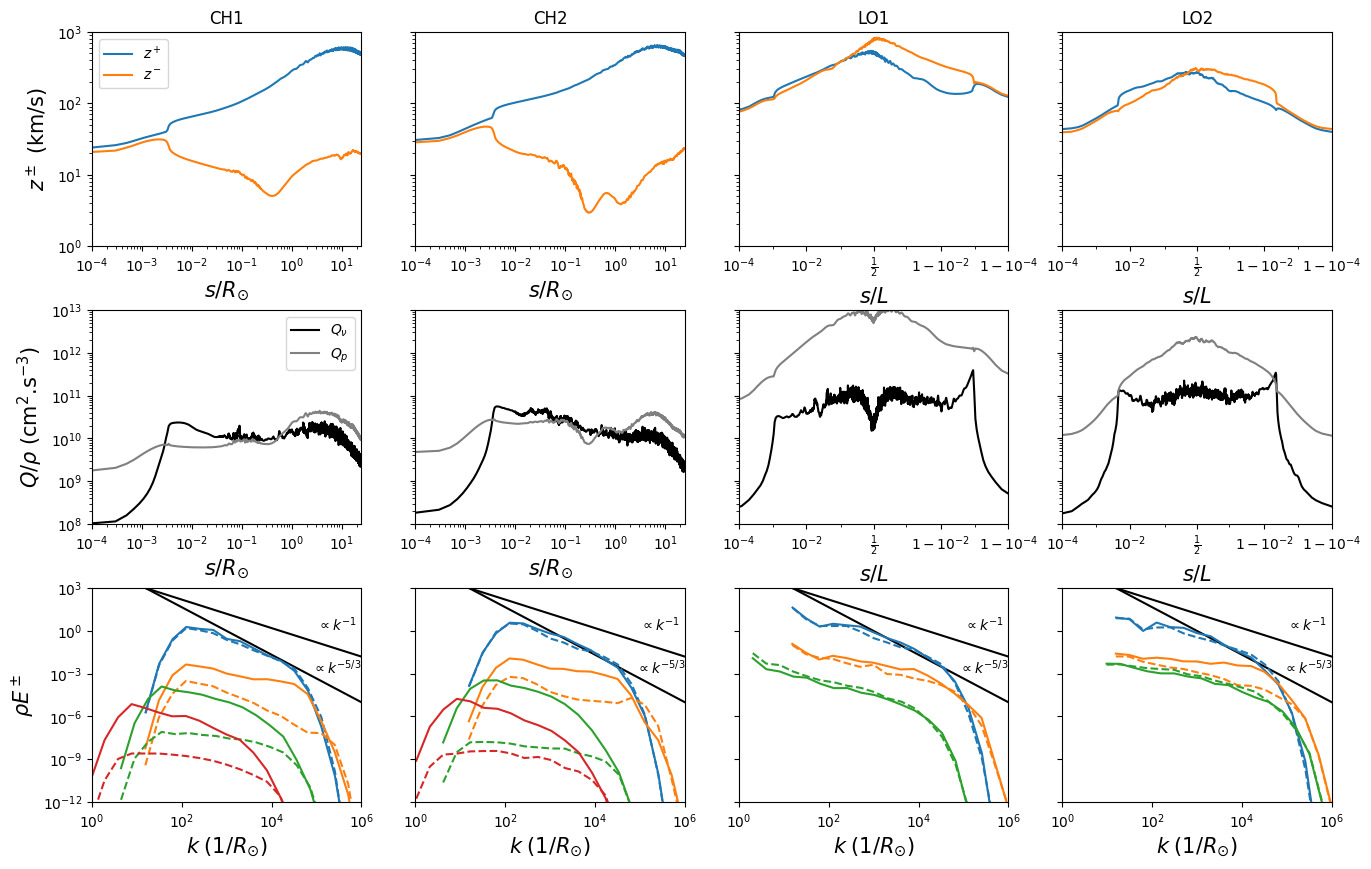} \\
\caption{Overview of shell simulations A with the same input parameters on the four geometries of coronal holes and loops. The top panel shows the time averaged rms fluctuations $z^\pm$. For coronal loop cases, the curvilinear abscissa $s$ is normalized by $L$ the length of the loop and both boundaries are shown in log scale. The middle panel shows the heating obtained by dissipation at small scales $Q_\nu$ in black and compares it with a phenomenological form $Q_p$. The last panel gives the spectra of the fluctuations ($+$ in plain lines, $-$ in dashed lines) at various distances of the Sun. For open regions simulations, the blue, orange, green and red lines corresponds to $r=1.001, 1.01, 2.5$ and $10 R_{\odot}$ respectively. For coronal loops, the blue and orange remain the same, while the green lines are the spectra at the apex of the loop.}
\label{fig:heating}
\end{figure*}

Table \ref{table:shell} sums up all simulations made with the shell model for this work. For each profile geometry, we explore three sets of turbulence parameters referred as cases A, B and C. In general, we chose the input parameters to create a turbulent heating close to what is observed in the (open) solar wind and as such used in the MHD described in section \ref{sec:tr}. We find that a transverse perturbation of $\delta v_{\odot} = 7$ km/s yields heating rate close to $10^5$ erg.cm$^{-2}$.s$^{-1}$ --- the minimum value to power the solar wind --- in both coronal holes, which defines our reference set of turbulent inputs: cases A. Cases B use a higher forcing amplitude $\delta v =10$ km/s, which logically results in a higher heating rate. Both cases A and B have the same injection scale $L^0_{\mathrm{inj}}= 35 000$ km, which is the largest scale introduced in the system and corresponds to the size of supergranules. Cases C are set with a injection scale ten times lower, close to the size of granules. To get the heating in the coronal holes at the right order of magnitude, we had to decrease the input velocity perturbations to $\delta v_{\odot} = 3$ km/s. All cases have the same correlation time $T_*=600$ s.

In Figure \ref{fig:heating}, we show some properties of the solutions of the shell model runs A for $\delta v_{\odot} = 7$ km/s, $T_\star=600$ s, and $L^0_{\mathrm{inj}} = 2\pi/(8 k_0) = 35000$ km. As the transverse perturbations are forced over three shells, the actual rms amplitude of the forcing is $\delta v_* = \sqrt{3 \times \delta v_{\odot}^2} \sim 12$ km/s. These parameters are close to the one used in \citet{Verdini2019} for a typical coronal hole solution. The results shown in Figure \ref{fig:heating} and in Table \ref{table:shell} are obtained with a time average of the solution between $150000$ and $200000$ seconds, which represents around ten Alfvén crossing times for the coronal holes profiles and more for the coronal loops. All simulations have reached a pseudo-steady state during the period. The top panel shows the outward and inward rms fluctuations $z^\pm$ as a function of the curvilinear abscissa $s$ along the profile. For coronal loops, we show $s/L$ in a logit scale, where $L$ is the total length of the loop, to emphasize the behaviour in the chromosphere and the transition region at both boundaries. 

In coronal holes simulations, one striking feature is the regime difference between a mostly balanced turbulence ($z^+ \sim z^-$) below the transition region and an imbalanced turbulence beyond. This can be seen also in the bottom panel of Figure \ref{fig:heating}, where the inward and outward spectra $E^\pm (k_n)= (z^\pm_n)^2$ are very comparable at $s=1.001 R_{\odot}$ (in blue), with a cascade covering four orders of magnitude and a spectral slope close to the usual Kolmogorov index $-5/3$. Higher in the corona, the outward wave dominates clearly while keeping a $-5/3$ slope, while the inward component has a flatter spectrum with slope close to $-1$. In coronal loops, both populations are everywhere well represented and spectra seems perhaps closer to $-1$ slopes. 
In the middle panel of Figure \ref{fig:heating}, we show the dissipation profile computed by the shell model. The plain black line is the true heating, while the grey line corresponds to the so-called phenomenological heating, a proxy often used in large scale fluid models \citep[see, e.g.][]{Dmitruk2002,VerdiniVelli2007,ChandranHollweg2009,Shoda2018a, Reville2020ApJS}. It can be written:

\begin{equation}
    Q_p/\rho = \frac{1}{2} \frac{|z^+|^2 |z^-| + |z^+| |z^-|^2}{2 L^0_{\mathrm{inj}}}.
\end{equation}
As said earlier, the total heating obtained in coronal holes solution $H_\nu$ is close to the input energy of the global MHD simulations, itself chosen to provide enough power in the solar wind (see Table \ref{table:shell}). Nevertheless, the total heating increases as we go from a typical polar coronal hole, to a slow denser wind around the streamer and to coronal loops. This is expected, and increased heating in the low corona is one explanation for the denser slower wind, which originates somewhere close to the loops. Interestingly, we see that the phenomenological heating is in general an overestimation of the true heating in the shell model \citep[as already noted in][]{Verdini2019}, except around the transition region. It does, however, seems to be a reasonable estimate in imbalanced turbulence region, i.e. in the open coronal wind regions. 

The properties of cases B are very similar to the one of cases A, only with larger wave amplitudes and larger heating rates. For cases C, the wave amplitude is logically smaller, and the cascade extent is also shorter as we remain with a fixed value of the viscosity/resistivity $\nu=\eta= 10^{11}$ cm$^2$.s$^{-1}$. In all cases, the dissipation starts to be significant for $k_n > 10^{4} R_{\odot}^{-1}$

\section{Ponderomotive force and FIP fractionation}
\label{sec:results}

\begin{figure}
\center
\includegraphics[width=3.3in]{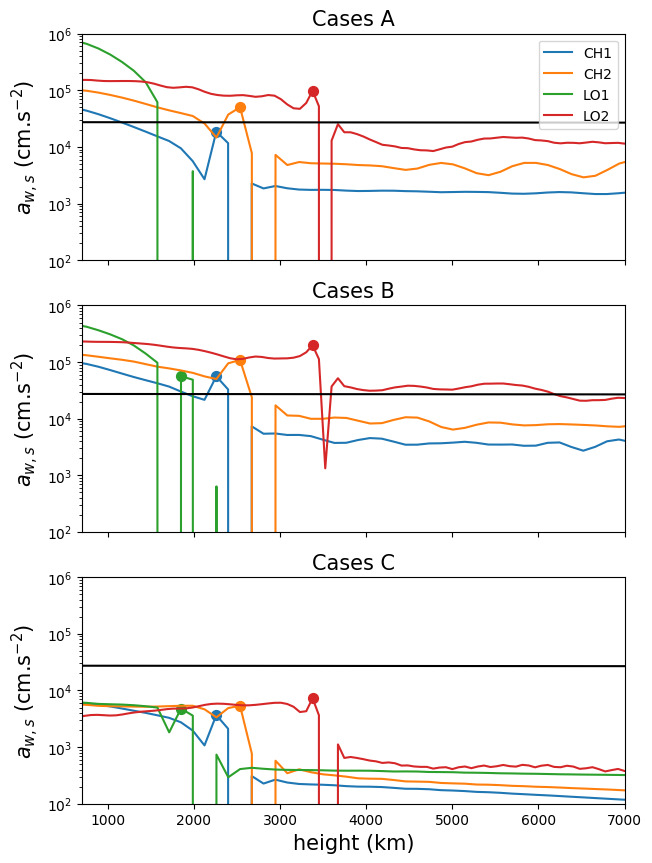}
\caption{Ponderomotive acceleration for all cases of Table \ref{table:shell} in the chromosphere and transition region. Top panel corresponds to cases A, middle panel to cases B, with a higher input $\delta v_{\odot}$, and bottom panel to cases C, with a lower amplitude and higher injection scale. The Sun's gravity pull is shown in black. The colored dots indicates the position of the transition region in the profiles. They are usually associated with a peak in the ponderomotive acceleration.}
\label{fig:pondforce}
\end{figure}

The main objective of this work is to compare the effect of the turbulence properties on the FIP fractionation for different background configurations. Among the key parameter of the current FIP models is the ponderomotive force, or the wave pressure exerted by the perturbations on the background flow. Figure \ref{fig:pondforce} shows the ponderomotive force obtained in all Table \ref{table:shell} cases, close to the inner boundary, in the chromosphere and transition region. Let us first study the cases A of Table \ref{table:shell}, shown in Figure \ref{fig:pondforce} in the top panel. The profile of $a_{w,s}$ has in all cases a similar shape, with a slow decrease and a peak located around the transition region. The TR peak, which is responsible for most of the FIP fractionation in the model that follows, has an interesting ordering. The LO2 profile, has usually the highest averaged ponderomotive acceleration and a higher peak, but the maximum of the LO1 is usually of the order of the coronal holes configuration peaks. Moreover, it seems that the amplitude of the peak is a growing function of the height of the TR.

For cases B, we observe for the ponderomotive acceleration essentially a shift up of all the curves, conserving the hierarchy of the previous cases as a function of the geometry. For most cases A and B, the strength of the ponderomotive force is higher than the opposing gravity of the Sun, around $\sim 2\times 10^4$ cm.s$^{-2}$, shown in black in Figure \ref{fig:pondforce}. The amplitude of the ponderomotive force is thus significant, especially at the transition region and Alfvén waves can have an influence on the coronal abundances. For cases C, however, the first striking feature is the net decrease of the ponderomotive acceleration below the Sun's gravity pull. This means that although the turbulent heating is perfectly compatible with what is necessary to power the solar wind (compare especially case CH1 A with CH1 C), this set of parameter will likely not create a low-FIP bias through the ponderomotive acceleration. In these last cases, the turbulence injection scale is 10 times smaller, i.e. the size of large granules. The initial amplitude is logically lower as the injection is made later in the cascade, which can explain the lower gradient of magnetic fluctuations. 

\begin{figure}
\center
\includegraphics[width=3.3in]{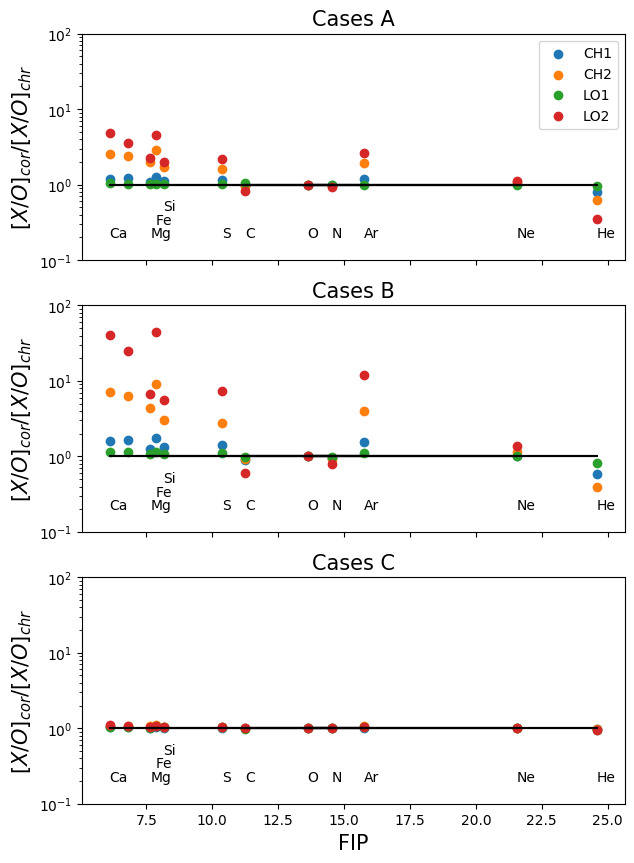}
\caption{FIP biases according to equation \eqref{eq:FIP} in the cases A (top panel), B (middle panel) and C (bottom panel) for usual minor ions. The integration is made between $z_{\mathrm{chr}}=700$ km and $z_{\mathrm{chr}}=7000$km above the solar surface. The turbulent velocity entering equation \eqref{eq:FIP} $v_T = 15$ km/s is constant.}
\label{fig:FIP1}
\end{figure}

We now compute the FIP fractionation created by the ponderomotive force in our models. We use for this the derivation of \citet{Laming2009}, that read: 
\begin{equation}
    \frac{\rho_j^{\mathrm{cor}}}{\rho_j^{\mathrm{chr}}} =  \exp \left( \int_{z_{\mathrm{chr}}}^{z_{\mathrm{cor}}} \frac{\xi_j a_{w,s} \nu_{\mathrm{eff}}}{\nu_{j,i} v_j^2} dz \right).
    \label{eq:FIP}
\end{equation}
Here, $\rho_j$ is the density of a given species $j$, $\xi_j$ the ionization fraction and 
\begin{equation}
    \nu_{\mathrm{eff}} = \frac{\nu_{j,i} \nu_{j,n}}{\xi_j \nu_{j,n} + (1-\xi_j) \nu_{j,i}},
\end{equation}
where $\nu_{j,i}$ and $\nu_{j,n}$ are the collision frequencies of the species $j$ with ions and neutrals respectively. We use the formulation of \citet{Schwadron1999} and \citet{Marsch1995} for the collision frequencies. The ionization fraction $\xi_j$ of each heavy ion is computed through an interpolation of Saha equilibria \citep{Saha1920,Saha1921} for $T \lesssim 10^4$ K and the CHIANTI database and the ChiantiPy interface \citep{Dere1997} for $T \gtrsim 10^4$ K. Finally, $v_j^2 = c_{s,j}^2+v_T^2$ is the (quadratic) sum of the thermal speed and of a mass independent turbulent velocity $v_T$, which represent a wave turbulence heating and is essential to avoid a mass dependant fractionation \citep[see][]{Schwadron1999}.

Figure \ref{fig:FIP1} shows the resulting FIP biases obtained with equation \eqref{eq:FIP} and a constant $v_T = 15$ km/s. We represent the minor ion density ratio between the low corona (7000 km above the photosphere) and the chromosphere (700 km above the photosphere), taking the Oxygen as a reference. The top and middle panel of Figure \ref{fig:FIP1} shows a clear low-FIP bias in cases A and B. Elements Fe, Mg, and Si, show bias up to a factor 10 or more compared with Oxygen, in the red case, i.e. the smallest coronal loop. The profile of the FIP bias follows the hierarchy of the ponderomotive force amplitude shown in Figure \ref{fig:pondforce}. For CH1 and LO1 cases, the low-FIP bias is weak and roughly similar, while it is generally higher for the CH2 configuration. Finally, as intuited, cases C do not show any clear FIP biases. This is a direct result of the much weaker ponderomotive force obtained with these simulation parameters, and the heavy ion densities are the same in coronal holes and coronal loops. 

\begin{figure}
\center
\includegraphics[width=3.3in]{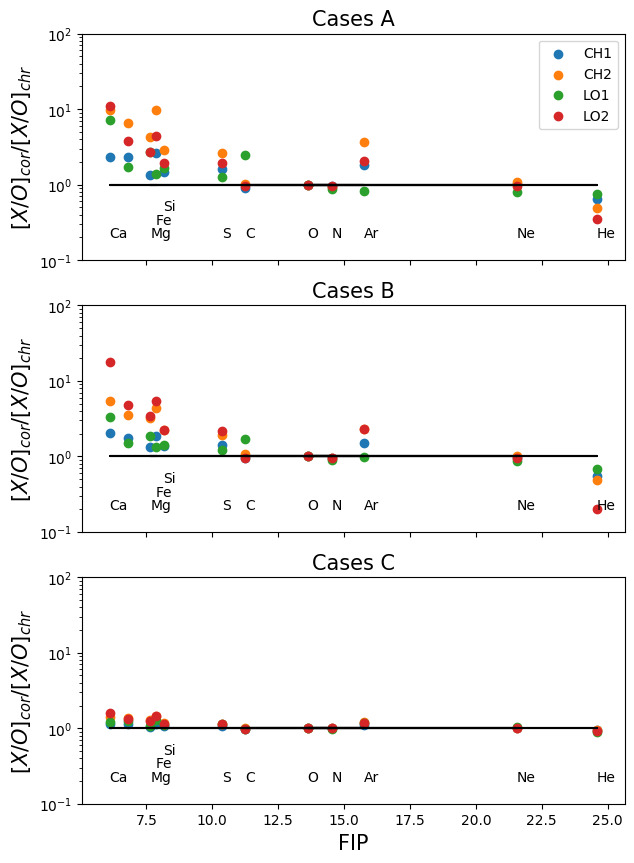}
\caption{Same as Figure \ref{fig:FIP1} but using $v_T$ as in equation \eqref{eq:vT}.}
\label{fig:FIP2}
\end{figure}

In Figure \ref{fig:FIP2}, we repeat the same exercise, but choosing a turbulent velocity
\begin{equation}
    v_T = \sqrt{\sum_{k_n > k_{\nu}} \delta v_n^2}.
    \label{eq:vT}
\end{equation}
We only sum the shell components for $k_n > k_{\nu} = 10^4 R_{\odot}^{-1}$, where dissipation is significant. The average amplitude of $v_T$ is around 8 km/s for cases A, 12 km/s for cases B, and 5 km/s for cases C. The choice of $v_T$ is crucial to get FIP bias comparable with observations. It needs to be large enough to avoid the gravitational settling, and thus the mass dependence of the abundances in the corona. However, if $v_T$ is too large, it will kill the low-FIP bias. The low-FIP bias of cases A is thus slightly enhanced, because the average turbulent velocity given by equation \eqref{eq:vT} is lower than the value used in Figure \ref{fig:FIP1}. The FIP bias of case B is quenched and cases C start to show enrichment in low-FIP elements. Moreover, the hierarchy between the different magnetic configurations is much less visible in Figure \ref{fig:FIP2}. CH2 cases have for instance, properties very similar to LO1 cases. This shows that a very careful treatment of the turbulence is important to build FIP fractionation models.

\section{Discussion}
\label{sec:ccl}

In this work, we have combined several models and tools to assess the efficiency of the Alfvénic turbulence in enriching abundances of low-FIP elements in the low corona. First, we used a global MHD model to get realistic profiles of the Alfvén speed and velocity gradients in both coronal holes and coronal loops, starting from the low chromosphere. They are indeed fundamental in the amplification, reflection and non linear interaction of propagating Alfvén waves. Then, we took advantage of the \shellatm\ code to compute the interaction between counter-propagating Alfvén waves, the turbulent cascade and the dissipation. Our approach certainly lacks self-consistency, as, for instance, the heating obtained with the shell model is significantly larger in coronal loops than in coronal holes (at least for cases A and B), while the MHD simulation has a much more homogeneous heating. It is, however, the first time that both coronal loops and coronal holes are treated with a shell turbulence model. This model solves the fully non-linear incompressible Alfvén wave equations, including the cascading process and the frequency dependant reflections and interactions of Alfvén wave populations. This is an improvement in comparison with analytical non-WKB approach of previous works \citep[see][]{Laming2004,Laming2009,Laming2012}. Moreover, the dissipation is treated physically at very high Lundquist numbers, which cannot be achieved in direct numerical simulations \citep[see, e.g.,][]{Dahlburg2016}.

A few comments on the applicability of incompressible MHD to this problem are in order. Reduced MHD equations are derived assuming small gradients in the guide field \citep[or even constant $B_0$, see][]{Strauss1976,ZankMatthaeus1992,ZankMatthaeus1993,Oughton2017} as well as in the density, conditions that are not verified in the upper chromosphere and the transition region. The excluded nonlinear couplings in the parallel and perpendicular directions (essentially compressible interactions) are responsible for the coupling of Alfv\'en waves to slow and fast magneto-acoustic modes, leading to steepening, shock formation and dissipation of slow magneto-acoustic modes and fast modes, and the refraction of the latter downwards back to the chromosphere for large perpendicular wavenumbers. The parametric decay instability (PDI), involving compressible processes, is also excluded. The latter plays a role in the formation of the turbulent spectrum and the balance of inward and outward Alfvén wave population, but is effective most in the lower beta regions higher up in the corona \citep[see][]{TeneraniVelli2013,Shoda2018b,Reville2018}. As noted in \citet{Verdini2019,Shoda2019}, density fluctuations created by the PDI increase the inward Alfvén waves amplitude and eventually increase the turbulent heating in the open wind regions.

Wave steepening, shocks and mode conversion, which do create a cascade in the direction parallel to the magnetic field, have typical spectra $\propto k^{-2}$ and are thus likely to be a secondary dissipation channel in comparison with the dissipation in the perpendicular plane, with spectra close to the Kolmogorov and Iroshnikov-Kraichnan phenomenology (see Figure \ref{fig:heating}). Alfvén wave coupling with magneto-acoustic modes may accelerate the perpendicular cascade, but because we run the simulation until a pseudo steady-state is reached, this is likely not a strong limitation of our approach. Numerous other modelers have made similar approximations to our own, as for example the studies by \citet{vanBallegooijen2011,PerezChandran2013,vanBallegooijenAsgariTarghi2016,vanBallegooijenAsgariTarghi2017,Chandran2019}. This study has no intention to be exhaustive, but we do examine a significant input parameter space in order to extract meaningful insights on the problem.

Our study shows that, assuming that turbulence is the dominant factor in the coronal heating and solar wind acceleration, a ponderomotive force can appear in the chromosphere and the transition region, and can be strong enough to create a low-FIP bias. This depends however on the turbulence parameters. Injecting energy at the scales of super granules provide the wave amplitude necessary for a low-FIP bias comparable with observations \citep[see, e.g.][]{Feldman1992}. The force is related to the a amplification of the waves in the corona and the strong gradient that appears at the transition region. Nevertheless, if the energy is injected at the scale of granules, the resulting ponderomotive acceleration seems too weak to explain the observations. Recent works have debated of the right injection scale parameters \citep{vanBallegooijenAsgariTarghi2017,Chandran2019}, and the amplitude of the ponderomotive force could be a way of constraining solar wind turbulence models. 

A second result of this study is that the low-FIP bias is not exclusive to coronal loops. Interestingly, we obtain significant low-FIP bias for open field configurations along the streamer (CH2). We also have usually little FIP bias for the LO1 loop, i.e. the loop at the very edge of the helmet streamer. This result might seem in opposition with previous works, such as the one of \citet{Laming2004,Laming2015}. Their result is indeed the consequence of resonant Alfvén waves in the loops, likely triggered by reconnection, which amplify significantly the ponderomotive acceleration. Injecting transverse motions from the inner boundary, we do not observe such resonances, and further studies are needed to compare the relative importance of different triggers. We do show however that resonances are not needed to obtain a low-FIP bias in coronal loops and open slow wind regions. This study thus questions, anyhow, the necessity of interchange reconnection to explain the composition of the slow solar wind.

Finally, it is important to stress that our modelling of the solar chromosphere is very simple. We assume, following \citet{Laming2004,Laming2009} and  writing equation \eqref{eq:FIP}, that proton drag, collisions and turbulent mixing, are exactly compensating for the Sun's gravity pull in the chromosphere. The actual balance of all these processes remains very hard to estimate. Shocks and compressible fluctuations are also believed to be significant in the chromosphere and even in the transition region \citep[see][for a recent review]{Carlsson2019} and are not accounted for in the shell simulations. Ongoing works are dedicated to build a compressible kinetic-fluid model, including heavy ions populations with various charge state and their interactions with protons and waves.

\section{acknowledgement}
This research was funded by the ERC SLOW{\_}\,SOURCE project (SLOW{\_}\,SOURCE - DLV-819189). We thank M. Laming for very useful discussions. Numerical computations were performed using GENCI grants A0080410133 and A0070410293. This study has made use of the NASA Astrophysics Data System.

\bibliographystyle{yahapj}
\bibliography{./biblio}

\end{document}